\documentclass[twocolumn, lengthcheck, amssymb, nobibnotes, aps, prd]{revtex4}
\usepackage{graphics}
\usepackage{graphicx}
\usepackage{ulem}

\usepackage{color}

\newcommand{\fN}{\tilde{N}}
\newcommand{\fo}{\tilde{o}}
\newcommand{\fh}{\tilde{h}}

\begin{document}

\title{Self-calibration of Networks of Gravitational Wave Detectors}
\author{Bernard F. Schutz}
\affiliation{School of Physics and Astronomy, Cardiff University, Cardiff, UK, CF24 3AA}
\affiliation{Max Planck Institute for Gravitational Physics (Albert Einstein Institute), 14476 Potsdam/Golm, Germany}
\email{SchutzBF@cardiff.ac.uk}
\author{B. S. Sathyaprakash}
\affiliation{Institute for Gravitation and the Cosmos, Pennsylvania State University, University Park, PA, 16802, USA}
\affiliation{Department of Physics, Pennsylvania State University, University Park, PA, 16802, USA}
\affiliation{Department of Astronomy \& Astrophysics, Pennsylvania State University, University Park, PA, 16802, USA}
\affiliation{School of Physics and Astronomy, Cardiff University, Cardiff, UK, CF24 3AA}
\email{bss25@psu.edu}

\begin{abstract}
Photometric calibration of astronomical telescopes normally involves using standard stars as reference objects; this is because a first-principles calibration based upon modelling the light path from the object to the photometer is extremely difficult. Gravitational wave detectors also need to be calibrated, and the standard method at present is  to use laboratory-based fiducial displacements that induce arm-length changes: electrostatic and/or laser light-pressure disturbances to the mirrors. As LIGO and Virgo improve their sensitivity and are eventually superseded by even more sensitive third-generation detectors, improving  calibration systems to keep pace with anticipated signal-to-noise improvements will be challenging. We explore here an alternative calibration method that uses astronomical signals, namely inspiral signals from compact-object binaries, and we show that it can in principle enable calibration at the sub-1\% accuracy levels needed for future gravitational wave science. This class of signal is a-priori well understood and modelled theoretically. We show how ensembles of these transient events can be used to measure the calibration errors of individual detectors in a network of three or more comparably sensitive instruments. As with telescopes, relative calibration of gravitational-wave detectors using detected events is easier to achieve than absolute calibration, which in principle would still need to be done with a hardware method for at least one detector at one frequency. Our proposed method uses the so-called null streams, the signal-free linear combinations of the outputs of the detectors that exist in any network with three or more differently oriented detectors. Signals do not appear in the null stream if the signal amplitude in the detector output is faithful to that of the real signal. Frequency-dependent calibration errors and relative calibration and timing errors between detectors leave a residual in the null stream. The amount of residual from each detector depends on the source direction. We adapt the method of matched filtering to the problem of extracting the calibration error of each detector from this residual. This requires combining linearly the filter outputs of a sufficient number of detected signals, and in principle it can achieve any desired accuracy in a long enough observation run. We anticipate that A+ detector networks, expected in 5 years, could employ this method to check anticipated hardware calibration accuracies. And with an expected harvest of several hundred binary signals per day, third-generation detectors should be able to control their calibration at the sub-percent level rapidly after any hardware change. Third-generation detectors may also be able to acquire high-accuracy calibration from the LISA mission, transferred via binary systems observed both in space and on the ground. And at some future time an ensemble of detected gravitational-wave pulsars could serve as a long-term memory for calibration, including for the absolute calibration. Such pulsars would be close analogues of optical astronomers' standard reference stars. 
\end{abstract}

\maketitle

\section{Introduction} 
\label{sec:intro}
\subsection{Calibration challenges}
\label{sec:challenges}
LIGO \cite{Science92,aLIGO_2015,Abbott_2020Scenarios} and Virgo \cite{AdV} have already changed our view of the Universe, among other things by uncovering a population of unexpectedly massive binary black holes like the first detection GW150914 \cite{Abbott_2016} and by enabling the intensive study of the kilonova that followed the merger of two neutron stars, GW170817 \cite{Abbott_2017}. When the most recent observing run O3 came to a close in March 2020, the detectors were already sensitive to four times more volume for such binary-coalescence signals as when GW150914 was detected. Further upgrades in sensitivity are planned through at least 2025, when the A+ hardware configuration \cite{Miller_2015} should become operational.

The accurate extraction of information from a gravitational wave signal that has a signal-to-noise ratio (SNR) equal to $\rho$ requires that the detector has been accurately calibrated at least to the level of $1/\rho$. And if there are systematic calibration errors shared by detectors in a network, then one should probably use the total network SNR in this expression. At present, LIGO estimates that the systematic error in calibration is at the 2\% level \cite{Sun_2020}. This is sufficient for all events detected so far, but it might become a limit on the accuracy of measurements made with ensembles of events, such as an accurate GW determination of the Hubble constant \cite{Schutz_1986, Chen_Fishbach_Holz_2018, Gray_2020,Bhattacharjee2020}. 

Calibration is difficult because the changes in separation of the mirrors inside interferometric detectors that gravitational waves produce are extraordinarily tiny. The current methods involve simulating signals by moving the mirrors artificially, either using electrostatic driving or photon pressure from a laser directed at the front of the mirror \cite{Karki:2016pht, Abbott_Abbott_2017,Bhattacharjee2020}. These are called fiducial displacements, and they have to be accurately modelled, because the disturbances they make are too small to be measured directly, independently of the interferometry. The accuracy of the modelling is improving, and in particular the photon calibrator itself can now be modelled to better than 0.5\% accuracy \cite{Bhattacharjee2020}. A technique currently under development uses the Newtonian gravity of a nearby spinning mass to produce the disturbance of the mirror \cite{Matone_2007,Estevez_2018,Inoue2018}. This also holds promise of going below 1\%, but much more work needs to be done. All these methods employ narrow-band fiducial displacements, which then have to be extrapolated to other frequencies across the observational band using a model of the detector response.

All astronomical observatories need to be calibrated, and  optical telescopes generally calibrate their photometric detectors using standard stars rather than artificial sources \cite{Deustua_Kent_Smith_2013}. The analogous approach for GW detectors would be to use strong continuous-wave signals that could be identified as standards (as mentioned in \cite{Pitkin_Messenger_Wright_2016}), so that a new detector or one that has undergone a hardware upgrade could re-calibrate itself to them. We return to this possibility at the end of this paper, but at present no such signals have been identified. The question we address here is, therefore, whether it is possible to use the detected {\it transient} signals from compact binary coalescence (CBC) events to accomplish calibration, and thereby provide an independent way to supplement and check whatever hardware calibration methods may be available. Transient signals are in a sense complementary to fiducial displacements, in that they are broadband disturbances that can in principle test the calibration of detectors across the observation band. They are suitable for testing calibration because they are well-modelled, so if general relativity is correct, then they can be used to discover calibration errors.

Calibration using transients is generally called {\it astrophysical calibration}, and there has been considerable interest in it in recent years 
\cite{Pitkin_Messenger_Wright_2016, Fairhurst_2018, Essick_Holz_2019,Payne2020}. Binary systems
Pitkin {et. al} \cite{Pitkin_Messenger_Wright_2016} discussed using a relatively strong signal to check whether the calibrated signal matched the model. They found that the calibration was consistent with the signal model. Later papers have investigated calibration errors within the context of matched filtering of weaker signals within networks of two or three detectors. Fairhurst \cite{Fairhurst_2018} considered how the timing information in a network of three detectors could identify calibration errors, while Essick \& Holz \cite{Essick_Holz_2019} simultaneously modeled both astrophysical parameters and calibration errors in multiple detectors as a way of constraining or measuring those errors under the assumption that GR is the correct theory of gravity and the astrophysical waveform family was perfectly known. In particular, they pointed out that it would be possible to accumulate such error measurements over successive event detections in order to make them more accurate, and that accuracies of 1\% could be achieved with 1000-2000 detected events. Only events that occur during a period in which the calibration is unchanged can be used in this way, and detector calibration teams typically adjust calibration a few times a year. Therefore, as Essick \& Holz remarked, requiring thousands of events for calibration makes the method unrealistic in current detectors. Two recent papers \cite{Payne2020,Vitale_2020} have used groups of detected events to test the calibration, using physically motivated calibration models. Both are consistent with the view that thousands of events will be needed to put strong constraints on these very small calibration errors.

We re-cast the astrophysical calibration problem in the framework of coherent detection in networks of three or more detectors, in which the detection of events and their timing at different detectors is part of a single solution. A network of $N$ detectors will have $N-2$ joint data streams, each of which contains in principle {\it only} the calibration errors. These are the so-called {\it null streams}, which we will discuss in detail below. The purpose of this paper is to show how to filter the errors out of a null stream, and then how to use the ensemble of all detected signals, not just the loudest ones, to extract the calibration errors of all the detectors in the network with sub-percent accuracy. We call this version of astrophysical calibration the {\it self-calibration} method. We believe that this is the optimum approach to using transient astrophysical signals for calibration. A given calibration epoch (a time-span in which the calibration is not changed) that lasts 4 months at A+ sensitivity might return 500 events, which we shall show below makes self-calibration potentially useful. And 3G detectors will return that many events in a few days.

It may at first seem impossible to use different events together to accomplish calibration. Each detected CBC signal differs from the others in the masses and spins of its components, and possibly also in the eccentricity of the orbits. Calibration requires something that is shared, some known reproducible information that can be used to detect calibration errors. What these CBC systems share is that the signals are well modelled theoretically across the spectrum, so that once a signal's parameters have been measured, any distortions of the apparent signal due to calibration errors are in principle identifiable by comparison with the theoretical model. This principle was the basis of the discussions in \cite{Pitkin_Messenger_Wright_2016, Fairhurst_2018, Essick_Holz_2019}, and we show how to take maximum advantage of it here.

Since a CBC source's distance is a-priori poorly known, self-calibration normally provides only a relative calibration, reducing frequency-dependent distortions but not correcting an overall amplitude error. The frequency dependence of calibration errors is difficult to model in present detectors. Such errors affect the parameters that we wish to extract from the signals, such as the masses and spins of component stars, and they affect the accuracy of many of the tests of general relativity that use these signals.  In a network, inconsistencies between detectors in the calibration of the amplitude across the frequency band (absolute calibration) produce mismatches that affect inferred sky position of sources and may therefore even inhibit identification of counterparts. All of these are addressed by self-calibration. But an overall error in absolute calibration shared by all detectors in the network produces an error in distances to binaries \cite{ Schutz_1986}, so self-calibration must generally be supplemented by some other method that can provide an absolute calibration of at least a single detector at a single frequency. The paper \cite{Essick_Holz_2019} looked to independent astronomical distance measures to provide the absolute information. This would, unfortunately, pass any systematic distance errors in the astronomical measurements on to the GW calibration. We prefer to hope that hardware calibration methods can be developed to the point where they can provide an absolute calibration at a few frequencies in at least one detector in the network, at whatever accuracy is required for the science. We will return to discuss absolute calibration in the final section.

Calibration errors have generally been modelled as systematic errors, affecting the data of all the events in a given calibration epoch in a similar way \cite{Sun_2020}. This is key to our method, as will be shown below. Errors are easier to model and measure if the calibration is a slowly varying function of frequency, but if not, then measuring the error using detected signals automatically weights the detected errors in different frequency bands according to the distortions they produce in the reconstruction of a signal. However, if calibration errors are correlated among detectors, something which we call network systematic errors, then we shall see that these tend to cancel out in our method and require more events to measure. This point was also made by Essick \& Holz \cite{Essick_Holz_2019}. And it may be that something in a detector's hardware changes during a given calibration epoch, changing the calibration error. We will come back later to how this could be monitored.

Self-calibration using very weak transient signals is made possible by the redundancy of the signal response among detectors in a network of three or more interferometers. If the direction to a source is known, and if the signal is observed by three comparably sensitive detectors, then (as we will review below) one can construct a linear combination of the three outputs that eliminates the two independent polarization waveforms and leaves just a combination of the instrumental noise. As mentioned earlier, this is called a {\it null stream}. It is constructed purely from the geometry of the detector network and the direction of the signal; the cancellation does not need to know the waveform of each polarization. First introduced by G\"ursel \& Tinto \cite{Gursel_Tinto_1989}, it was given its name by Wen \& Schutz \cite{Wen_Schutz_2005}, who explored how it can be used to veto instrumental artefacts that masquerade as signals. Chatterji et al \cite{Chatterji_2006} developed these ideas further, as did Wen \cite{Wen_2008}, who also showed how the null stream could be used as a coherent detection statistic. The mathematics of null streams and of extracting the desired information from a system of redundant outputs with independent noise is fully explored in the papers just referred to.

Signals cancel in the null stream if they are represented faithfully in the outputs of the three detectors. Calibration errors distort this representation, leaving residuals in the null stream. These residuals are extremely small: the signals are already below broad-band noise, and the errors are 2\% or less of the signal amplitude. We show that they can nevertheless be measured by constructing matched filters that make use of the fact that the calibration error of each detector is a systematic error, a property encoded in the residual from each detected signal. By accumulating the output of an appropriately constructed matched filter linearly over many events, the systematic becomes measurable. We estimate that thousands of events will be needed, which makes the method useful for third-generation detectors, which are expected to make hundreds of CBC detections each day. These detectors will require sub-percent-level calibration because they will make many detections with SNR higher than 100.

\subsection{Principles of the self-calibration method}\label{sec:principles}
Here we focus on the way calibration errors appear in the null stream. In order to take advantage of the redundancy of the signal, the detector outputs have to be expressed in terms of strain, or gravitational wave amplitude $h$, before they are combined. This conversion to $h$ depends on understanding the calibration. The calibration is expressed as a parametrized function in the frequency domain that whose parameter values are adjusted to best fit the calibration as measured across the whole signal spectrum. Any errors in calibration can propagate into the null stream to cause incomplete cancellations of gravitational wave amplitudes, as was noted by Chatterji et al \cite{Chatterji_2006}. But if the signal waveform is a-priori well understood, and if its parameters are well-measured by the network, then the residual signal in the null stream will be the product (in the frequency domain) of the calibration error and known weighted amounts of signal. 

In order to detect this residual against the detector noise of the null stream with as much accuracy as possible, one has to apply a matched filter in much the same way as the signal itself is detected in the data stream of each detector. As we will explain in Sec.~\ref{sec:calibration}, the calibration error function will be parametrized in some way in the frequency domain, so the aim is to find best-fit values of these parameters. This gives a family of functions over which one searches to find the best fits to the residuals in the null stream. By adding the SNR output of the matched filters over a number of detected events, eventually the best-fit filter will stand out. In this way, the network calibrates itself, using the known waveform. 

Naturally, calibration can only be done over the frequency range spanned by the signal itself. Binary inspirals sweep across the range from the lowest detectable frequency to their coalescence frequency. Binary neutron stars are ideal because they go up to beyond 1~kHz. Binary black hole signals terminate at lower frequencies, but will still be useful because calibration errors are typically larger at lower frequencies.  

A couple of cautions should be noted. First, for global networks of detectors, the direction to the source of the signal must be known in order to form the correct 
%linear combinations, because in order to make the waveforms cancel coherently between different detector output streams, they must be shifted in time according to the arrival times at the different detectors. 
matched filters, since the amount of signal from any one detector that is present in the null stream depends on the location of the source on the detector's sky. A three-detector network already provides this direction information \cite{Wen_Schutz_2005} with enough accuracy to allow the null stream to be constructed, so independent identifications of sources is not necessary. Second, the signals being used for calibration must all arrive before there is any change in the hardware of any of the detectors that could require re-calibration. As noted in \cite{Essick_Holz_2019}, the large number of events needed (estimated below) makes it unlikely that self-calibration can be useful for current detectors, but it may become useful with LIGO's A+ sensitivity. And for 3G detectors like the Einstein Telescope \cite{Maggiore_2020} or Cosmic Explorer \cite{Abbott_Abbott_Abbott_2017}, the event rate should be more than adequate.

%For the Einstein Telescope (ET), self-calibration is even easier, in that it is not necessary to know the direction to the source [ref?] (see Sec.~\ref{sec:ET}). The three independent interferometers in the ET design can be treated as being co-located, because their separations are very much smaller than the wavelengths of signals in their detection band. This means that no time-shifts are needed, so any one detector's signal output is a simple linear combination of those of the other two, independently of the direction to the source: the cancellation is simply a matter of the geometry of the detectors' layout. 

We turn in the next two sections to describe the way self-calibration works. 

\section{Definition of the null stream}\label{sec:definition}
The null stream of three detectors is a consequence of the fact that, in general relativity, a gravitational wave has only two polarization amplitudes, which act 
transversely to the direction of propagation. If we adopt a coordinate system $(\theta,\,\phi)$  for the celestial sphere and at any point on the sphere define an orientation for local $x$-$y$ coordinates in the plane tangent to the sphere, then we can resolve a wave coming from that direction into its amplitudes $h_+(t)$ and $h_\times(t)$. An interferometric detector located on Earth will respond to the two wave amplitudes linearly with response functions $F_+$ and $F_\times$ that depend on the direction to the source, the location of the detector, its orientation with respect to local North, and the normal direction to the plane formed by its two arms. A more detailed discussion can be found in many references \cite{Thorne_1987, Sathyaprakash_Schutz_2009, Andersson_2020}. For our purposes we don't need to write down the explicit expressions for these antenna pattern functions. 

We assume that we have an array of three detectors, no two of which are  both co-aligned and co-located, {\it i.e.} that any two responses contain linearly independent combinations of the two polarizations. The outputs $o(t)$ of the detectors (calibrated to give the detector strain) contain not just the strain response to the incoming wave but also detector noise $n(t)$. The three outputs can be written as follows, in terms of the wave amplitudes at time $t$ at detector number 1 and the delays in arrival $\tau^{(2)}$ and $\tau^{(3)}$ at detectors number 2 and 3, respectively: 

 % \textcolor{blue}{BSS: Is it not better to use a notation in which the wave arrives at the geo-center at time $t$ and at the three detectors at times $t+\tau_k,$ where $\tau_k=\vec d_k\cdot \hat n/c,$ where $\vec d_k$ is the position vector of the detector and $\hat n$ is the direction to the source? This can make the equations look more symmetric.}
 % \textcolor{red}{Answer from BFS: Well, that makes ET more complicated, no? We don't know the arrival direction, so we don't know the geo-center arrival time.}
%
\begin{widetext}
\begin{eqnarray}
o^{(1)}(t)&=&F_{+}^{(1)}h_{+}(t)+F_{\times}^{(1)}h_{\times}(t)+n^{(1)}(t)\label{h1}\\
o^{(2)}(t+\tau^{(2)})&=&F_{+}^{(2)}h_{+}(t)+F_{\times}^{(2)}h_{\times}(t)+n^{(2)}(t+\tau^{(2)})\label{h2}\\
o^{(3)}(t+\tau^{(3)})&=&F_{+}^{(3)}h_{+}(t)+F_{\times}^{(3)}h_{\times}(t)+n^{(3)}(t+\tau^{(3)})\label{h3}.
\end{eqnarray}
\end{widetext}
This equation assumes that the direction to the source is correctly known, so that the time-delays and the values of $F_+$ and $F_\times$ for each detector can be correctly calculated. It also assumes that the signal duration is short, so that the position of the source on the sky does not change measurably during the detection. This keeps $F_+$ and $F_\times$ independent of time. Of course our analysis could be generalized to the case of long-duration signals.  

Any two of these equations can be solved for $h_{+}(t)$ and $h_{\times}(t)$, and the results can be substituted into the third equation, giving a linear function of the three detector outputs in which the gravitational wave amplitudes do not appear. This linear combination of appropriately time-shifted and weighted detector outputs is the null stream $N^{123}(t)$ of this set of three detectors: 
\begin{equation}\label{nullstream}
N^{123}(t):=A^{23}o^{(1)}(t)+A^{31}o^{(2)}(t+\tau^{(2)})+A^{12}o^{(3)}(t+\tau^{(3)}),
\end{equation}
where
\begin{equation}\label{determinant}
A^{ab}=(F_{+}^{(a)}F_{\times}^{(b)}-F_{+}^{(b)}F_{\times}^{(a)}).
\end{equation}
In the ideal case, where there are no errors of calibration and where one is using the correct direction to the source, the signal content of the output streams cancel and one has a pure noise null stream that we call $N^{123}_\mathrm{n}$. If we introduce a couple of  convenient changes in notation, 
\begin{equation} \label{notation}
A^{(1)} := A^{23}, \quad A^{(2)} := A^{31}, \quad A^{(3)} := A^{12}, \quad \tau^{(1)} := 0,
\end{equation}
then we can write the null stream noise compactly as 
\begin{equation}\label{inullstream}
N^{123}_\mathrm{n} = \sum_a A^{(a)}n^{(a)}(t+\tau^{(a)}).
\end{equation}

We shall assume from now on that the noise is normally distributed with zero mean. Note that the function of the time-delays is to effectively shift the detectors into a plane parallel to the incoming wavefronts. If one has three differently oriented detectors already in the same location, they are in the same wavefront plane no matter what direction the wave is coming from, so no time-delays are needed. In this case the cancellation happens for signals coming from {\it any} direction. This applies to the proposed ET detector, and also for LISA. We discuss these cases in Sec.~\ref{sec:ET} below. 

%It also applies in part to the LISA space-based gravitational wave detector, now under development by ESA [ref]. LISA will be a triangular array like ET, but the interferometers centered on each vertex of the triangle are not fully independent: they share arms, whereas in ET they have parallel arms in separate optical systems. LISA forms a Sagnac-type interferometer, and the null stream is called the Sagnac mode. At low frequencies, the signal cancels from the Sagnac mode, as it does in ET. But LISA's frequency band extends to frequencies high enough that their wavelengths are shorter than the arm-length, and in that case the vertices are not normally all in the plane of the wavefront, and the null-stream cancellation does not happen without appropriate time-shifts. 

%Sharing arms in LISA means that the noise content of the outputs of the different interferometers are not independent, as they are in ET for the most part and in arrays of separated interferometers. In our discussion below we will often make use of this statistical independence, so our results do not transfer to the LISA case.

The null stream has a number of uses. One \cite{Wen_2008} is to localize a source on the sky: one computes the null stream for different angular positions of the source on the sky, and the true position is the one where no evidence of signal remains. Another \cite{Wen_Schutz_2005} is to use it to recognize and veto non-Gaussian noise events in one or more detectors that masquerade as signals. Unlike true incoming signals, such ``glitches'' do not disappear in $N^{123}$. Our interest here is to use the null stream to identify and correct calibration errors. 

\section{Calibration and the null stream}\label{sec:calibration}
% Sathya - can you start out with a short description of calibration. We should define c^a(f) in terms of the calibration function (true one and the one that is used).
In this Section we will briefly discuss how the interferometer response is calibrated and then use that procedure to motivate the definition of complex calibration functions $c^a(f)$ we use. In the following sections we will demonstrate how the null stream formed using the response function of a network of detectors can be used to self-calibrate gravitational-wave detectors. 

\subsection{Calibration of interferometers}

The calibration of interferometers is a complex and delicate process. The value of the gravitational-wave induced strain $h$ has to be inferred from the output of the photodiodes at the destructive-interference port of the interfeometer. But account must be taken of feedback systems used to control the positions of the mirrors, which ensure that the interferometer remains locked. All of these systems need to be modelled and then their parameters measured, to form an overall model of how the gravitational-wave strain amplitude $h$ can be inferred from the measured output. This has been described in detail for current LIGO interferometers in a number of papers \cite{Tuyenbayev_2017, Sun_2020,Cahillane2017}.

Although the detector is calibrated in the time-domain (in order to generate the correct $h(t)$ to be recorded in the output signal stream), the sensing and actuation functions are modeled in the Fourier domain, from which one constructs the desired digital filters for application in the time-domain \cite{Viets_2018}. The calibration system -- modelling the measurements, feedback, and standard signal generators -- contains many parameters, each of which could have a measurement error.

As mentioned earlier, the calibration model is tested and its parameters measured regularly by applying fiducial displacements to the interferometers, in which mirrors are moved by electrostatic and photon-pressure forces \cite{Tuyenbayev_2017}. Since the resulting motions are too small to measure in any way other than by looking at the output of the interferometer, these tests require good models of the forcing mechanisms and of how the mirrors respond to them. Any errors in these models or in the measurements associated with them will result in errors in the calibration that is applied to the signal output to get $h$. 

Calibration is not expected to change much during an observation run, that is over timescales of weeks and months. 
%Can we use the figures Sathya generated??
The errors therefore feed into the output as systematic errors, affecting every signal in the same way. We shall assume in this paper that self-calibration is a way to measure this constant error in each detector in a network.

The calibration error is a complex function of frequency across the detector's observation band. This is estimated by the detector calibration team and provided to the data analysis teams as a fractional systematic error in the signal $\fh(f)$ in the Fourier domain. In what follows we shall call this the {\it fractional error function} $c(f)$. We shall not be concerned with the (hidden) parameters of the calibration system. We seek only to measure $c(f)$. The interpretation of this measurement and its use for correcting the models of the calibration system and of the fiducial displacements is then left to the calibration teams. 

\subsection{Null-stream calibration of interferometers}
\subsubsection{Null stream in the frequency domain}

As we have just seen, calibration of interferometers is applied in the frequency domain, so we shall work with the Fourier transform of the null stream:
\begin{equation}\label{fnullstream}
\fN^{123}(f)  := \sum_a A^{(a)}\fo^{(a)}(f)e^{2\pi i f\tau^{(a)}}. 
\end{equation}
If there is a fractional calibration error $c(f)$, then the $h$-content of the null stream in the frequency domain will contain $c(f) [F_{+}\fh_{+}(f)+F_{\times}\fh_{\times}(f)]$ from each detector, weighted with the antenna pattern determinants:
\begin{equation}\label{fcalerrors}
\fN^{123} = \fN_\mathrm{n}^{123} + \sum_a A^{(a)} c^{(a)}(f) [F_{+}^{(a)}\fh_{+}(f)+F_{\times}^{(a)}\fh_{\times}(f)]e^{2\pi i f\tau^{a}},
\end{equation}
where the sum is over the detectors, and where $c^{(a)}(f)$ are complex functions that account for error in both amplitude and phase of the signal. This equation contains both the null-stream noise $\fN_\mathrm{n}^{123}$ and what we will call the {\it null-stream calibration error signal}, 
\begin{equation}\label{errorsignal}
\epsilon^{123}(f) := \sum_a A^{(a)}c^{(a)}(f) [F_{+}^{(a)}\fh_{+}(f)+F_{\times}^{(a)}\fh_{\times}(f)]e^{2\pi i f\tau^{a}}.
\end{equation}

In Eq.~\ref{fcalerrors}, if we know the signal waveforms $\fh_{+}(f)$ and $\fh_{\times}(f)$, then the only unknowns are the calibration error functions. The aim of self-calibration is to use estimates of the null-stream calibration error signal of a set of detected signal events to determine the error functions $\{c^{(a)}(f)\}$.

To see what this involves, let us make a further simplifying (but very commonly valid) assumption that there is no precession of the plane of the binary during the observation, so that the polarization can be taken to be constant in time. In this case both $\fh_+$ and $\fh_\times$ are proportional to a single signal model that we shall simply call $\fh(f)$. We assume that the parameters of the signal have been measured, as has the polarization, so that we can write
\begin{equation}\label{components}
\fh_+(f) = \alpha \fh(f), \qquad \fh_\times(f) =\beta \fh(f),
\end{equation}
where $\alpha$ and $\beta$ are complex constants, independent of frequency, and the same for all detectors. This assumption simplifies the contribution of each detector to the null stream, so that Eq.~\ref{errorsignal} becomes
\begin{equation}\label{errorsignalsimple}
\epsilon^{123}(f) = \sum_a [\chi^{(a)}(f)c^{(a)}(f)]\fh(f),
\end{equation}
where we define the complex functions 
\begin{equation}\label{chidef}
\chi^{(a)}(f) := A^{(a)}[\alpha F_+^{(a)} +\beta F_\times^{(a)}]e^{2\pi i f\tau^{a}}.
\end{equation}
We shall refer to these coefficients as  {\it geometry parameters} because they are known functions of the geometry of the network and source. Then the null stream Eq.~(\ref{fcalerrors}) becomes
\begin{equation}\label{fcalerrors2}
\fN^{123} = \fN_\mathrm{n}^{123} + \sum_a [\chi^{(a)}(f)c^{(a)}(f)]\fh(f). 
\end{equation}
 The sum in this equation is the signal we are looking for.

\subsubsection{Matched filters for the null stream: an overview}
In order to have detected the signal in the first place, the signal streams of all three detectors had to be matched-filtered using a family of templates of the signal \cite{Sathyaprakash_Schutz_2009, Jaranowski_Krolak_2009, Creighton_Anderson_2011}. This is because the signal itself is normally too weak to be recognized in broad-band noise. It follows from this that the residual we seek in the null stream will be even further below the broadband noise, because we are dealing with errors at the 2\% level or smaller. So the best way to extract as much of the desired information as possible about the error signal $\sum_a [\chi^{(a)}(f)c^{(a)}(f)]\fh(f)$, buried underneath the null stream noise $\fN_\mathrm{n}^{123}$ in Eq.~(\ref{fcalerrors2}), is to apply matched filtering to it as well. 

Although matched filtering is a standard technique in GW detection, its application to this problem has some novel features. We describe them first before we go on to the algebraic implementation, in order to make the method more transparent.
The aim is to determine the calibration errors $c^{(a)}(f)$, which will be approximated by functions that depend on a set of parameters. Matched filtering should extract the best set of values of these parameters. The residual error in the null stream in Eq.~(\ref{fcalerrors2}) contains the functions $c^{(a)}(f)$ we seek to determine, but we can't filter just for them because they are multiplied by $\chi^{(a)}(f)\fh(f)$. Our filters must therefore be constructed to model the entire residual. Since the detector network has already measured the signal, we know $\chi^{(a)}(f)\fh(f)$, so the correct filter family is straightforward to construct by just multiplying this by the parametrized approximation to $c^{(a)}(f)$, and then applying filters for various sets of parameters that cover the parameter space, just as one does when searching for CBC signals.

However, even for the best-match filter, the expectation of the filter output from an event will be far below the filter noise. This is because, as inspection of the residual in Eq.~(\ref{fcalerrors2}) will show, the size of the residual is roughly the size of the original event $\fh$ times the calibration error. Therefore, if the error $c^{(a)}(f)$ is only of order 2\%, then the network SNR of the filter output for the residual will be of order 2\% of the original SNR of the detected event. If the event had a network SNR of 20, then the expected SNR of the output of the {\it best-fit} residual filter applied to the null stream will be of order 0.4. 

This requires us to combine filter outputs from different events in order to build up a detectable SNR for the calibration, in much the same way as Essick \&  Holz did \cite{Essick_Holz_2019}. Since each event will have a different function $\chi^{(a)}(f)\fh(f)$, {\it the family of matched filters changes for each event}. Nevertheless, by adding linearly the SNR of the filters labelled with the same set of parameter values applied to each successive event, we will build up SNR in the usual Gaussian way, proportional to the square-root of the number of events. If we set a detection threshold of SNR~$=5$ for this cumulative measure, and we start with an expectation of 0.4, then we need about 150 events, each detected with network SNR of 20. This may well be possible for an observing run with A+ technology, where events may be expected more than once a day, and it should be easy to reach in only a few days for the planned 3G detectors, which should be able to measure sub-percent errors in a several-month observation run. 

It is important to realize that a measurement of $c(f)$ for each detector at the 5-sigma level allows the calibration not just to be checked, but to be improved, by adjusting the calibration model parameters, the model itself, or the fiducial displacement systems. It is not unreasonable to expect that in this way a 2\% calibration error can be pushed down to around 1\%. A 1\% calibration error could itself then be measured and reduced using another 600 events, but that might be too many for an A+ run to detect before a hardware upgrade changes the calibration model.

\subsubsection{Construction of the matched filters for transient events}
Now we turn to the algebraic implementation of the scheme we have just described. We address this problem in two steps. In this part we construct the appropriate filters for the null stream of a single event. In the following part we construct the optimal filter for the problem of using multiple events to find the best description of the calibration errors of the detectors.

To filter optimally one needs to model the individual calibration error functions with template families that depend on parameters, creating a parametrized set of possible error functions that can approximate the true error $c^{(a)}(f)$ accurately. This has to be done by the experiment teams that look after the calibration of each instrument. Here we simply denote the parametrized calibration error function of the $a^\mathrm{th}$ detector as  $C^{(a)}(\vec{p}^{\,(a)};f)$, where $\vec{p}^{\,(a)}$ is the set of $n^{(a)}$ parameters used to model the calibration error of this detector. These do not necessarily correspond to any parameters that the calibration error function $c^{(a)}(f)$ might depend on, but the intention is that for some (``best'') parameter set $\vec{p}^{\,(a)}$ the functions $C^{(a)}(\vec{p}^{\,(a)};f)$ will be good approximations to the true errors $c^{(a)}(f)$, so that finding these parameter values (with appropriate uncertainties) will guide the calibration teams in readjusting the calibration.

Among the parameters in the parameter set $\vec{p}^{\,(a)}$ of the $a^\mathrm{th}$ detector will be, besides parameters that describe the calibration error, also the time-shift $\tau_{1a}$ that enters the construction of the null stream in Eq.~\ref{nullstream} and the geometry parameters $\chi^{(a)}$ in Eq.~\ref{chidef}. (The time-shift $\tau_{11}$ vanishes identically since we have adopted the convention that the shifts are referenced to the signal arrival time at detector 1.) If there are timing errors at detector sites (differences between local clock time and reference UTC), then these will put signal power into the null stream. These network timing errors were first discussed in this context by Fairhurst \cite{Fairhurst_2018}.
 
The parametrized family of filters for finding the calibration error signals is found by replacing $c^{(a)}(f)$ in $\epsilon^{123}(f)$ with $C^{(a)}(\vec{p}^{\,(a)};f)$. Each filter depends on all three parameter sets. We call these filters $E^{123}(\{\vec{p}\};f)$, where $\{\vec{p}\}$ is a shorthand that stands for the set $\{\vec{p}^{\,(1)},\,\vec{p}^{\,(2)},\,\vec{p}^{\,(3)}\}$:
\begin{equation}\label{fullfilter}
E^{123}(\{\vec{p}\});f) := \sum_a [\chi^{(a)}(f)C^{(a)}(\vec{p}^{\,(a)};f)]\fh(f).
\end{equation}

To filter the null stream Eq.~\ref{fcalerrors2} optimally \cite{Thorne_1987, Sathyaprakash_Schutz_2009}, we must weight the filter integrals inversely using the null-stream  spectral noise density, which is given by
\begin{equation}\label{sh}
S_h^{123}(f) = (A^{23})^2 S_h^{(1)}(f) + (A^{31})^2 S_h^{(2)}(f) + (A^{12})^2 S_h^{(3)}(f).
\end{equation}
Here we use the notation $S_h^{(a)}(f)$ for the spectral noise density of detector $a$. The null-stream spectral noise density is just this weighted sum of the spectral noise densities of the individual detector streams (no cross-terms) because we assume their noise distributions are independent on one another. (This would not be true for the case of LISA, as we mentioned above.) The spectral noise density defines an inner product on the space of possible signals \cite{Sathyaprakash_Schutz_2009} in the frequency domain:
\begin{equation}\label{innerproduct}
\left <x,y \right > := \int_{-\infty}^{\infty}\frac{x(f)y^*(f)}{S_h^{123}(f)}df.
\end{equation}

Using this notation, the application of the filter in Eq.~\ref{fullfilter} to the data stream Eq.~\ref{fcalerrors2}
gives the output
\begin{equation}\label{filteroutput}
\Phi(\{\vec{p}\}) =  \left <\fN^{123},E^{123}(\{\vec{p}\}) \right >.      
\end{equation}
This will contain a noise component from  $\fN_\mathrm{n}^{123}$ (the noise content of the null stream $\fN^{123}$) and a deterministic part that is the weighted correlation of the filter for these parameters with the calibration residual in the null stream. As long as the detectors' noise is Gaussian and zero mean (which we assume throughout), the filter output $\Phi(\{\vec{p}\})$ will be normally distributed with mean $\phi(\{\vec{p}\})$ equal to its expectation value, 
\begin{widetext}
\begin{eqnarray}\label{filtermean}
\phi(\{\vec{p}\}) &:=& \overline{\Phi(\{\vec{p}\}) } = \left <  \epsilon^{123},\,E^{123}   \right > 
= \int_{-\infty}^\infty \frac{ \sum_{a,b} \chi^{(a)}(f) c^{(a)}(f) [\chi^{(b)}(f) C^{(b)}(\vec{p}^{\,(b)};f)]^*|\fh(f)|^2}
{S_h^{123}(f)}df,
\end{eqnarray}
and variance equal to 
\begin{equation}\label{filtervariance}
\mathrm{Var}(\{\vec{p}\}) := \left <E^{123}(\{\vec{p}\}),E^{123}(\{\vec{p}\})\right> =  \int_{-\infty}^\infty \frac{ |\sum_a \chi^{(a)}(f) C^{(a)}(\vec{p}^{\,(a)};f)|^2\,|\fh(f)|^2}
{S_h^{123}(f)}df.
\end{equation}
\end{widetext}
Therefore the signal-to-noise ratio  (SNR) for the output of a given filter is 
\begin{equation}\label{snrevent}
\rho(\{\vec{p}\}) = \frac{\Phi(\{\vec{p}\})} {[\mathrm{Var}(\{\vec{p}\})]^{1/2}}.
\end{equation}

The calibration error models  $C^{(a)}(\vec{p}^{\,(a)};f)$ for each detector must be computed for a sufficiently dense set of sampled parameter values $\vec{p}^{\,(a)}$ that covers the entire volume of the family in parameter space. We call this family the {\it calibration error template bank}. The filters $E^{123}(\{\vec{p}\};f)$ are formed by multiplying each member of the calibration error template bank by the known signal waveform and geometry parameters in frequency space. 

The outputs $\phi(\{\vec{p}\})$ from these filters will of course be dominated in each individual event by the null stream noise, because as remarked above the calibration error is a small fraction of the signal itself. Therefore it will be necessary to apply these filters to a statistically sufficient number of detected events in order to find the best-match filter and therefore the best estimate of the null stream calibration error signal $\epsilon^{123}(f)$ as defined in Eq.~(\ref{errorsignal}). We turn now to show how this is done in an optimal way. 

\subsubsection{Cumulative matched filtering}
The different signal detection events are assumed to occur in an observation span during which the detector calibration model is unchanged and the detectors are operating normally. We also assume that the different signals do not overlap with one another in time, so that the null stream can be divided into sections, each of which contains a single signal, and whose noise is statistically independent of the noise in each of the other sections. Since detector noise is colored, the low-frequency noise below the signal must be filtered in order to achieve this statistical independence. (The assumption that signals do not overlap will need care in 3G detectors, and will probably require that only events stronger than a certain threshold should be selected for self-calibration.) Of course, the signal waveform $\fh(f)$ and the geometry parameters $\chi^{(a)}(f)$ will change from event to event. But we want to apply the matched filters for any given parameter set $\{\vec{p}\}$ to all these events, and use their outputs to decide with parameter set is optimum. This choice is therefore a property of the {\it ensemble} of all the events that are being used. 

%\textcolor{red}{Sathya, this paragraph follows your excellent heuristic reasoning. Can we derive it in a Bayesian way??}
This requirement can be framed as a form of matched filtering, extending the familiar approach to finding a single signal in a data stream. The data set in this case is the {\it union} of the non-overlapping segments of the null stream that contain the different events. The problem is to find the best filter for the detecting the signal -- the calibration error -- using the information in the whole union. If we were looking for a single signal, we would simply multiply the filter by the data in Fourier space and add the products up (integrate) over all frequencies, inversely weighted by the noise spectral density. We do this because the noise at each frequency is assumed uncorrelated with that at other frequencies (a property of stationary noise). The best filter is the one that maximises the integral. In our present case, the noise at each frequency in each of the data sets in the union can be taken to be independent, so by extension we construct, for each parameter set $\{\vec{p}\}$, a filter that is the union of the filters $E(\{\vec{p}\};f)$ given in Eq.~\ref{fullfilter}, and we sum the product of this filter with the data in Fourier space over frequency {\it and} event number. We shall return in the next section to use maximum likelihood to derive the detection statistic, but first we give a more heuristic derivation that might make it clearer.

To sum over events, we must address the normalization of the filters. The sum can be thought of as a one-dimensional random walk, in which each step has a random element (null-stream noise) and a deterministic value (the expectation value of the filter output). The best parameter set $\{\vec{p}\}$ is the one that produces the longest walk through the accumulation of consistent values of the deterministic part. The optimum random walk for discovering the correct parameter values is one in which the noise distribution is the same at each step. Then the expected random displacement increases as the square-root of the number of steps, while the deterministic parts accumulate quasi-linearly. This means that instead of adding up the output values $\Phi(\{\vec{p}\})$ we must add up the SNR values $\rho$ as given in Eq.~\ref{snrevent}.

Suppose we use $N$ detection events for self-calibration. If we introduce now a label for each event -- using capital Roman letters $A$, $B$, $C$, \ldots -- then the full detection statistic $\rho_\mathrm{SC}$ for self-calibration by the member of the calibration error bank with the parameter set  $\{\vec{p}\}$ is 
\begin{equation}\label{detectionstatistic}
\rho_\mathrm{SC}(\{\vec{p}\}) = \frac{1}{\sqrt{N}}\sum_{A=1}^N \rho_A(\{\vec{p}\}) = \frac{1}{\sqrt{N}} \sum_{A=1}^N\frac{\Phi_A(\{\vec{p}\})}{\left[\mathrm{Var}_A(\{\vec{p}\}) \right]^{1/2}}. 
\end{equation}
This statistic follows the normal distribution with variance 1, so its value can be regarded as the SNR for the given member of the calibration error bank over the entire set of observed events. The best set of parameters $\{\vec{p}\}$ for representing the calibration error is the set that maximizes $\rho_\mathrm{SC}$.

There will be three main reasons that the SNR of a given filter will differ from one event to another, apart from statistical fluctuations: (i) the event's amplitude $|\fh(f)|$ can be louder or quieter; (ii) the null-stream noise $S_h^{123}(f)$ can change either because the weighting of the different detectors changes with the incoming direction of the signal or because of non-stationarity in the detectors; and (iii) the geometry parameters $\chi^{(a)}$ change because of the change in the incoming direction of the signal. By examining Eq.~\ref{filtermean} and Eq.~\ref{filtervariance}, one can see that $\rho_A$ for event $A$ scales linearly with a scale increase in the amplitude of the event or an increase in the geometry factors, so that stronger events contribute more to the detection statistic than weaker ones. Similarly, a scale increase in $S_h^{123}$ produces a decrease in $\rho_A$ by the square-root of the scale, in other words $\rho_A$ is inversely proportional to the noise amplitude during the event. So again, the contribution of noisier events is down-weighted.

As noted earlier, this method cannot by itself determine the overall absolute calibration, although amplitude will be one of the parameters for each detector's error model, and this will allow the absolute calibration of the various detectors at least to be coordinated. Signal events that have identified counterparts with independently determined distances can be used for this, if the distances are regarded as reliable \cite{Essick_Holz_2019}. But it would be more satisfactory to obtain the absolute calibration, at a minimum of one frequency in one detector, using a hardware method, as we remarked earlier. Then self-calibration can be compared with the hardware calibration at other frequencies, as a check on the hardware method.

Once found, the best-fit parameter values should be used to re-calibrate the detectors, and then to re-determine the parameters of the detected signals, such as component masses and spins, and sky locations and polarizations. If these change significantly from the values that were used as input into the self-calibration procedure, then a further iteration will be needed in order to get the best calibration. If the calibration errors initially were small, then one can hope that only one such iteration, at most, should be necessary.

\subsubsection{Matched filter statistic as Bayesian likelihood}
We can interpret the detection statistic in Eq.\,(\ref{detectionstatistic}) as Bayesian likelihood function. To this end we will assume that the each detector output is normally distributed with zero mean. The noise at different frequencies is further assumed to be uncorrelated, which is equivalent to saying that in the time-domain the noise is stationary, and the variance at frequency $f$ is simply the noise PSD $S_h(f).$ This can be a good assumption at most times, and therefore around most detected events.

The null stream described in Eqs.(\ref{nullstream}) and (\ref{fnullstream}), being linear combinations of the data from different detectors, will continue to be stationary and Gaussian. We will first address filtering applied to the null for individual events, and then combine the events to get the linear sum statistic Eq.\,(\ref{detectionstatistic}). For clarity in this section we introduce some simplifying notation. The null stream for event $A$ will be denoted $ \mathbf{x}_A := \tilde N^{123}_A(f),$ the expected calibration signal for that event will be denoted $ \mathbf{y}_A:= E^{123}_A(\{\vec p\,\}, f)$ and the background noise during that event by $ \mathbf{n}_A := \tilde N_{n,A}^{123}(f).$

If the calibration applied to the data is perfect then we do not expect any residual and in that case $ \mathbf{x}_A= \mathbf{n}_A.$ On the other hand, inaccurate calibration will leave behind a residual and in that case $ \mathbf{x}_A= \mathbf{n}_A +  \mathbf{y}_A.$ In the absence of any residual the probability of getting data $ \mathbf{x}_A$ is the same as probability of getting a realization of noise. Since $ \mathbf{n}_A$ is a Gaussian with {\bf zero} mean we have 
\begin{equation}
P( \mathbf{x}_A |  \mathbf{n}_A) \propto \exp\left [ -\frac{1}{2} \left <  \mathbf{x}_A,  \mathbf{x}_A \right > \right ],
\end{equation}
where the inner product of a pair of vectors $ \mathbf{x}_A$ and $ \mathbf{y}_A$ is defined in Eq.\,(\ref{innerproduct}). If the calibration is not perfect then $ \mathbf{x}_A$ contains residual calibration and hence the mean of $ \mathbf{x}_A\ne 0.$ However, $ \mathbf{x}_A- \mathbf{y}_A$ will obey a normal distribution with zero mean and hence 
\begin{equation}
P( \mathbf{x}_A |  \mathbf{y}_A) \propto \exp\left [ -\frac{1}{2} \left <  \mathbf{x}_A- \mathbf{y}_A,  \mathbf{x}_A - \mathbf{y}_A \right > \right ].
\end{equation}
The likelihood function for the filtering of event $A$ is the ratio of the two probabilities we computed above:
\begin{eqnarray}
\Lambda_A &:=& \frac{P( \mathbf{x}_A |  \mathbf{y}_A)}{P( \mathbf{x}_A |  \mathbf{n}_A)}\nonumber \\
&\;=& \exp\left [ -\frac{1}{2} \left <  \mathbf{x}_A- \mathbf{y}_A,  \mathbf{x}_A - \mathbf{y}_A \right > + \frac{1}{2} \left <  \mathbf{x}_A, \mathbf{x}_A \right > \right ].
\end{eqnarray}

Exploiting the linearity of the inner product we see that the likelihood function for event $A$ reduces to:
\begin{equation}
\Lambda_A = \exp\left [ \left <  \mathbf{x}_A,  \mathbf{y}_A \right > - \frac{1}{2} \left <  \mathbf{y}_A, \mathbf{y}_A \right > \right ].
\end{equation}
If this likelihood ratio could be significantly larger than 1, then we could stop here and use the usual matched-filtering detection statistic for a single event, namely that the logarithm of the likelihood should be a maximum. This would lead as usual to the parameter set that most closely approximates the true calibration error at the time of that event, which would maximize of the SNR $\rho_A$ as given in Eq.\,(\ref{snrevent})

However, as we have seen, individual events will all be expected to have likelihood ratios very much smaller than 1, and so the best filter parameters can only be determined by accumulating likelihood over events. Since by assumption the null streams of different events are statistically independent, the full self-calibration likelihood $\Lambda_\mathrm{SC}$ over the whole set of $N$ events is simply the product of the likelihoods of the individual events:
\begin{eqnarray}\label{totallikelihood}
\Lambda_\mathrm{SC} &:=& \prod_{A=1}^N \Lambda_A\nonumber \\
&\;=& \exp \left \{ \sum_{A=1}^N \left [ \left <  \mathbf{x}_A,  \mathbf{y}_A \right > - \frac{1}{2} \left <  \mathbf{y}_A, \mathbf{y}_A \right > \right ] \right \}.
\end{eqnarray}
Maximizing the logarithm of this quantity leads precisely to the detection statistic Eq.\,(\ref{detectionstatistic}), apart from the convenient (but unimportant) normalizing factor of $1/\sqrt{N}$.

\section{Application of null-stream calibration}
\subsection{Globally distributed networks of detectors}
For the next decade or more, any application of self-calibration will involve detectors that are distributed around the Earth, with non-zero time-delays and mostly uncorrelated antenna patterns. In this section we discuss in a qualitative way how the method might work for various subsets of three detectors. 

First, let us make a rough estimate of the number of detected events that are needed to correct the calibration errors. Suppose that the fractional calibration error is of order $\delta \ll 1$. Then for the parameter set $\{\vec{p}\}$ for which the element $C^{(a)}(\vec{p}^{\,(a)})$ of the calibration error bank matches the corresponding calibration error $c^{(a)}(f)$ for each of the detectors, Eq.~\ref{filtermean} and Eq.~\ref{filtervariance} suggest that the expectation $\phi_A(\{\vec{p}\})$ will be of order $s\delta$, where $s$ is the (linear) network SNR of the detected event itself. In order for the detection statistic to have an expected value of, say, 5, then one needs of order $(5/s\delta)^2$ events. If we have $\delta = 0.02$  (a 5\% calibration error) and if our events themselves have SNR of $s=20$, then (as we estimated before) we need around 150 events. 

This could be an underestimate since we have not taken account of the significant ``look-elsewhere'' effect, which is that the number of elements in the calibration error bank will be very large, and so there may be several random peaks at $\rho_\mathrm{SC} = 5$. But this effect drops rapidly as we raise our threshold, so we are optimistic that the number of events required to determine the correct filter will not be much larger than 150, for the above parameters.  

We have done this estimate for a relatively strong SNR=20 event. There will be many weaker ones as well. Should one use only a few strong events rather than the whole detected population? For a given detector network, the number of events stronger than a given threshold SNR is proportional to the inverse cube of SNR, since it depends only on the volume of space that can be surveyed. The number of events needed to reach a given desired detection statistic is, as we have just seen, proportional to the inverse square of the event SNR. So as one raises the threshold for selecting events for self-calibration, the number of events available decreases faster than the number required to maintain the detection statistic. The lesson is that one should use all reliably detected events. This may well offset the ``look-elsewhere'' effect just mentioned, so that a run that has 150 events above SNR=20 may well be sufficient to measure a 2\% calibration error if all the weaker detections are used as well. Numerical simulations with realistic numbers of error parameters will be needed to verify this.

It seems unlikely that a single observation run of current detectors (LIGO, Virgo, and KAGRA) \cite{aLIGO_2015, AdV, KAGRA} in the next year or two will be able to return 150 such events during a calibration epoch lasting a few months, even if Virgo and especially KAGRA improve their sensitivities relative to that of LIGO, which itself is still improving. Having four detectors means that there are two independent null streams, which should help a little. But when KAGRA reaches a  sensitivity comparable to that of the other detectors, this will improve the sky coverage of the network and also the amount of time when at least three comparable detectors are observing in coincidence, which could increase the number of events usable for self-calibration by a factor of 2 or 3 \cite{Schutz_2011}.

The near-alignment of the two LIGO detectors means that null streams involving both will be most sensitive to calibration errors in those two detectors. If the calibration errors of the two LIGO detectors have a significant correlation with one another, essentially because they use the same calibration methods, then this correlated part might be difficult to find in the null stream. That is because the determinants $A^{23}$ and $A^{31}$ will typically have similar magnitude but opposite sign (note the ordering of their indices), and this will lead to partial cancellation of any correlated errors. However, in practice the calibration errors of the two LIGO detectors do not seem to be well correlated \cite{Sun_2020}, because other factors are more important, such as hardware differences that affect parameters in the calibration model.

The next step in sensitivity will be LIGO's A+ upgrade \cite{Miller_2015}, which has already been funded. Moreover, LIGO-India \cite{Indigo,Unnikrishnan:2013qwa} is under construction and plans to operate from the beginning with A+ sensitivity. At this point, having 150 events in a given calibration epoch, detected just with the LIGO detectors, seems possible. The Virgo and KAGRA detectors can only make things better. For this reason we see the $A^+$ era as the the start of the real application case for self-calibration. 

The issue of the correlation of calibration errors between detectors could be even more problematic for the three-site LIGO network, where the entire network might be calibrated by the same model. Inspection of Eq.~\ref{nullstream} shows that if all three detectors have the {\it same} calibration error, then the calibration errors cancel out completely, regardless of where the event is on the sky, and even for separated detectors. The reason is that the null stream cancellation depends on the presence of the same wave amplitudes $h_+$ and $h_\times$ in all detectors. If these are replaced by mis-calibrated amplitudes $(1+\epsilon)h_+$ and $(1+\epsilon)h_\times$ in all detectors, with the same $\epsilon$, then they will again cancel completely. When LIGO-India comes online, if the calibration model is shared among all LIGO detectors, there is therefore the possibility that shared systematic errors will not be found by the self-calibration method.

%\textcolor{red}{Sathya - by a rough calculation I think that the accuracy we can reach is given by the single-detection SNR divided by sqrt(n-detections). But this could be off, especially when there are a lot of parameters. In fact, it could be that we replace n by n divided by the number of parameters to be determined. Can you independently think about this please?}

%\textcolor{blue}{Bernard - I think it is OK to scale by $n$ as long as they all have the same SNR and we can defend this from the Fisher matrix calculation. The problem is that we will have far more weaker events than louder ones but it is the louder ones that will likely influence the calibration parameters. Therefore, I am not sure we can simply divide by the number of events but weight the number of events by SNR.}

\subsection{Sky-independent null stream}\label{sec:ET}
The design of the proposed 3G detector ET envisages three V-shaped interferometers, one each at the three vertices of an equilateral triangle. The sum of the responses of the three interferometers, as we shall see below, is a null stream no matter where the source is in the sky. In fact, this is true more generally for any configuration that has a closed topology. Consequently, self-calibration with ET is significantly simpler.

We denote by $h^a(t),$ $a=1,\dots,3$ the response functions of the three detectors in the ET array. By definition, $h^a=F^a_+ h_+ +F^a_\times h_\times,$ where as before $h_+$ and $h_\times$ are the plus and cross polarizations of the incident signal and the same for the three detectors. The antenna pattern functions $F_+, F_\times$ depend on the direction to the source $(\theta,\phi)$ and the polarization angle $\psi$ and they are given in terms of the detector tensor $d^A_{ij}$ and polarization tensors $e_{+,\times}^{ij}$ \cite{Sathyaprakash_Schutz_2009}:
\begin{equation}
    F^a_{+} = \sum_{ij} d^a_{ij} e_+^{ij}, \quad  F^a_{\times} = \sum_{ij} d^a_{ij} e_\times^{ij}.
\end{equation}
If $\mathbf{e}_1,$ $\mathbf{e}_2$ and $\mathbf{e}_3$ are unit vectors along the three arms of the triangle (taken in a cyclic order) then the detector tensors are given by 
\begin{eqnarray}
    d^1_{ij} & = & \frac{1}{2}\left ( \mathbf{e}_2^i \mathbf{e}_3^j - \mathbf{e}_3^i \mathbf{e}_2^j, \right ),\\
    d^2_{ij} & = & \frac{1}{2}\left ( \mathbf{e}_3^i \mathbf{e}_1^j - \mathbf{e}_1^i \mathbf{e}_3^j, \right ),\\
    d^3_{ij} & = & \frac{1}{2}\left ( \mathbf{e}_1^i \mathbf{e}_2^j - \mathbf{e}_2^i \mathbf{e}_1^j, \right ).
\end{eqnarray}
Plugging in the expressions for the detector tensors in the equation for the antenna pattern functions above one can explicitly see that $\sum_a h^a=0.$ Thus, the null stream of ET is just the simple sum of the response functions, and is called the \emph{Sagnac combination.} 

The null stream therefore does not have to be re-constituted for each event. Neverthless, the filters used do depend on the location of the event in the sky, since they involve the parameters $\chi^{(a)}$, which weight the amount by which any one detector's calibration error affects a particular event. So the formalism above is substantially unchanged. 

ET will have the same issue as three-detector LIGO, in that shared systematic calibration errors will cancel in its null stream. Given the promise of self-calibration during the 3G detector era, every effort needs to be made to avoid shared systematic calibration errors among the detectors.

LISA also has a Sagnac combination, for the same reason as ET: it is a triangle of three interferometers. But LISA differs from ET in two important respects. First, the signals may have wavelength comparable to or shorter than the arm-length, so that the response functions are more complicated, and such signals do not cancel in the Sagnac combination. Second, and more important, in LISA each interferometer based on a vertex of the LISA array shares a common laser link with the others, so that only two of the three interferometers have independent noise distributions. So the assumptions that we have made to arrive at self-calibration do not apply in the case of LISA. LISA can in fact calibrate itself in other ways, using the wavelength of its laser light as an on-board standard. 

\section{Discussion and Future prospects}
Calibration of detectors is a tricky business, involving accurate modelling of the mechanisms that move mirrors in calibration tests. Having independent ways of checking these models is important, and self-calibration provides a way of doing this that is completely independent of laboratory mechanisms. 

\subsection{Higher-accuracy calibration}
As we have seen, self-calibration can, with a sufficient number of events, calibrate to accuracies well below that needed for individual event parameter estimation.  This is important, because some of the science will come from ensembles of detected events. Measuring the Hubble Constant with standard sirens, for example, could be done to an accuracy of 2\% in the next few years \cite{Chen_2018}, and this would be sufficient to contribute significantly to the current puzzle that $H_0$ measurements seem to group around two distinct values. But for such a measurement to be reliable, calibration has to contribute a smaller error than this, so 1\% or better is desirable. This is because calibration errors can be systematic: they won't average down in the way that the measurement errors for $H_0$ do. We have seen that measurements in the A+ era, if they still have the 2\% calibration errors of current measurements, could be corrected to the 1\% level by self-calibration, using many of the same events that would be used to measure $H_0$.

% Mention systematic errors, which we have ignored. Selection effects in the choice of events; presence of glitches in the null stream at a very low amplitude; systematic errors in parameter measurements from event to event, such as clock errors that could locate the sky position wrong.

\subsection{Calibration prospects in the 3G/LISA era}
As gravitational wave networks get more sensitive, calibration become an even more pressing problem. In the 3G era, detectors like ET will be detecting several binary coalescence events per minute, and the science that can come from using large ensembles of events is exciting: accurately measuring $H_0$ and other cosmological parameters \cite{Maggiore_2020}, calibrating electromagnetic distance-ladder populations \cite{Gupta_2019}, searching for cosmic anisotropy, using weak lensing and microlensing of signals to explore cosmic structure \cite{Liao_2019, Hannuksela_2020}, testing general relativity in ultra-strong gravity, and so on.  These may need calibration accuracies at the sub-percent level. Self-calibration should be able to reach these levels, given the enormous detection rate in the 3G era, but with one big caveat: it cannot measure an overall mis-calibration of amplitude. This is a special case of the shared systematic calibration error we have discussed. 

If hardware calibration methods in the 3G era are able to provide sufficiently accurate absolute calibration -- even in just one detector at one frequency -- then the problem is solved, and self-calibration can be used to measure calibration errors between detectors and across the frequency band of the incoming signals. More than likely, such a method will be used for all detectors at a variety of frequencies, and then self-calibration will at least provide a useful consistency check on the frequency-dependence of the calibration in each detector. And if the event rate is high enough, as is likely in the 3G era, time-dependent calibration errors (due to detector changes within a given calibration epoch) can be searched for by separately combining events detected in different stretches of time.

Another intriguing possibility for obtaining an absolute calibration is to transfer calibration from the LISA mission to the detectors on the ground, while LISA is operating (launch 2034, operation for up to ten years). This is possible because there will be a special class of binary systems that are detected by both LISA and the ground-based networks \cite{Sesana_2016}. Hundreds to thousands of such events may be detected if LISA manages a 10-year operation span. LISA will, as we remarked, do its own on-board calibration, which is also referred to as self-calibration, but which does not use event waveforms. And it will be very accurate, potentially better than $10^{-4}$. When the same system is detected in its final coalescence phase by the ground-based detectors, months to years later, its amplitude will already be known to high accuracy by extrapolating the waveform (with LISA's measured parameters) to ground-based frequencies. (This requires that antenna pattern effects can be compensated for in order to get the true amplitude, which in turn requires a good sky position. The LISA observations should be able to provide this for most binaries.) To transfer the absolute calibration and nothing else, one uses the mechanism of self-calibration described above, but restricts it to finding just the overall absolute calibration. It should be possible to use only one filter: the one that is already known to be optimum for the relative calibration of the network. This would allow the detection threshold for measuring the absolute calibration to be reduced to 3, since there is no confusion over filter parameter values. If absolute calibration is desired at the 1\% level, and if typical events have a network SNR of 20, then one needs about 225 events during a calibration epoch of the ground-based network.  

The LISA mission will have a finite lifetime, so the question arises of whether it might be possible to preserve a ``memory'' of the LISA calibration that would be useful to ground-based detectors after the end of the mission.  This leads us back to what, at the beginning of this paper, we described as the ideal astrophysical calibration signal: a long-lived continuous signal from a gravitational wave pulsar. We have focused in this paper on short-duration binary merger signals, because that is the only kind of signal that has so far been detected. But there are intensive ongoing searches for GW pulsar signals in LIGO data \cite{Abbott_Abbott_2019}. It is very possible that many would already be detectable if their locations were known, but a blind all-sky search to LIGO's sensitivity limit is still computationally impossible \cite{Papa_2016}.  But it may well be that by the time LISA flies in 2034, through a combination of improved computer power and improved ground-based detector sensitivity, an ensemble of tens or even hundreds of GW pulsars will be known. Their signals are weak, but when observed over long periods of time, filtered and added together in the way we have added short-duration events together, it is well possible that they could allow calibration at the sub-percent level. Such an ensemble could preserve any absolute calibration inherited from LISA or from a ground-based hardware method, allowing self-calibration thereafter to be a completely independent way of checking hardware calibration methods.

\section*{Acknowledgements}
We are indebted to Madeline Wade, Jeff Kissel, and Richard Savage for discussions on detector calibration and helping us to understand the current process of calibration of LIGO data and to Madeline Wade and Chun-Fung Wong for a careful reading of the manuscript.  BSS and BFS gratefully acknowledge support from the Science and Technology Facilities Council (STFC) of the United Kingdom. BSS is supported in part by NSF grants PHY-1836779, AST-1716394 and AST-1708146.  

%\textcolor{red}{More acknowledgements to calibration team, plus check that each reference has a page number.}

\bibliographystyle{plain}
\bibliography{references}

\end{document}